\documentclass[preprint,preprintnumbers]{revtex4}
\usepackage{amssymb}
\usepackage{latexsym}
\usepackage{amsmath}
\usepackage{graphicx}% Include figure files
\usepackage{bm}% bold math
\usepackage{setspace}
\makeatletter
\newenvironment{figurehere}
  {\def\@captype{figure}}
  {}

\newenvironment{tablehere}
 {\def\@captype{table}}
 {}

\makeatother
\begin{document}
\bibliographystyle{apsrev}
\title{Conductance of a Conjugated Molecule with Carbon Nanotube Contacts}
\author{Nicolas A. Bruque}
\email{nbruque@ee.ucr.edu}
\author{M. K. Ashraf}
\author{Roger K. Lake}
\affiliation{Department of Electrical Engineering, University of
California Riverside, CA 92521}
\begin{abstract}
\noindent
Calculations of the conductance of a carbon nanotube (CNT)-molecule-CNT
structure are in agreement with experimental measurements
\cite{Nuckolls_Sci06}. The features in the transmission correspond
directly to the features of the isolated molecular orbitals.
The HOMO provides conductance at low bias that is relatively
insensitive to the end groups of the cut CNTs, the cut angle,
or the number of molecular bridges.
A molecular conformation change not directly in the path of the
carrier transport increases the resistance by over 2 orders of
magnitude.
\\
\\
{\em Keywords}: Electron transport, F{\footnotesize IREBALL}, CNT,
DFT, NEGF, conductance, molecular electronics
\end{abstract}
\maketitle
%
%
%\section*{Introduction}
Individual molecules have been proposed as the ultimately scaled
electronic device in future electronics. Gold metal contacts to
molecules has been studied the longest and given the most attention
both experimentally and theoretically
\cite{Reed_Sc97,Taylor_rectification_PRL02,DAMLE:CP:2002,Ratner_Interface_geom_NL05}.
The agreement between the experimentally measured currents and the
theoretically predicted currents in these systems was
difficult to obtain due to the strong dependence
of the conductance on the contact geometry
%and molecular conformation
\cite{Lindsay_Ratner_Adv_Mat07,Kaun_Seideman_PRB08,Quek_NanoLett07,Strange_Jacobsen_JChemPhys08,Ratner_Interface_geom_NL05,Tomfohr_Sankey_J.Chem.Phys04}.
The correspondence has greatly improved with the
introduction of amine linkers
\cite{Venkataraman_amine_Nat06,Venkataraman_amine_NL06,Kosov_amine_contacts_PRB07}.

The carbon nanotube (CNT) has been
utilized as an alternative contact to single molecules
\cite{Nuckolls_NL07,Nuckolls_Sci06,He_Tour_NatMat06}.
%To build such structures, a gap in the CNT $<$ 10 nm is cut using
%ultrahigh-resolution electron-beam lithography followed by an oxygen
%plasma etch. The CNT is functionalized at the interface with
%carboxylic acid allowing the molecules with amide end groups to
%chemically attach across the gap.
The CNT-molecule-CNT amide
interface chemistry provides a well defined covalent bond to attach
CNT contacts to single molecules. The CNT contacts can provide
both metallic and semiconducting properties governed by the CNT
chirality.

Prior transport studies report that the current response of
CNT-molecule-CNT systems is greatly influenced by the chirality of
the CNT contacts
\cite{Qian_CNT_mol_CNT07,Ke_CNT_mol_PRL07,Cuniberti_NatureNano07}
while others report changes when examining the passivation chemistry
at the cut ends of the CNTs \cite{Guo_CNT_mol_CNT07}.
The CNT-molecule-CNT transport studies
published to date have been limited to model or proposed systems
\cite{Pandey_PSSA_MolRTD06,Bruque_PRB1,Akdim_JPCC_08,Qian_CNT_mol_CNT07,Ke_CNT_mol_PRL07,Cuniberti_NatureNano07,Guo_CNT_mol_CNT07,Qian_CNT_2BDT_PLA07}
where the CNT-molecule interface chemistry, interface geometry, CNT
chirality, or molecular conformation were the areas of focus.
Comparisons between theory and experiment for these systems have not
yet been performed. In this paper we present transport calculations
of a CNT-molecule-CNT system that was built and measured
\cite{Nuckolls_Sci06}, we make quantitative comparisons, and
we show
how the features in the transmission spectrum result from the features
of the molecular orbitals of the isolated molecules.

%\section{Approach}
To calculate the equilibrium transmission of a CNT-molecule-CNT system, our method
uses density functional theory (DFT) implemented by the \emph{ab
initio} tight-binding molecular dynamics code F{\footnotesize
IREBALL} \cite{Sankey89,Further_developments,Jelinek_prb_05} coupled
with a non-equilibrium Greeen's functional (NEGF) algorithm
\cite{Seminario_NEGF01,Xue_JChemPhys01,DAMLE:PRB:2001,XUE:CP:2002,Bruque_PRB1}.
The BLYP exchange-correlation functional is used to perform a
self-consistent calculation
\cite{Demkov_charge_transfer95,Approx_dens_fn98} using a double
numeric $sp^{3}$ localized orbital F{\footnotesize IREBALL} basis.
We relax the system using periodic boundary conditions until
forces are $<$ 0.05 eV $\AA^{-1}$ using a self-consistent convergence
factor of $10^{-5}$.
These matrix elements are used
to calculate the surface self-energies, Green's function
of the device, and the resulting transmission. To calculate
the room
temperature conductance, we take the derivative of the current
equation with respect to voltage,
$G = \frac{2e}{h} \int dE T(E)(-\frac{\partial f}{\partial E})$,
where $f$ is the Fermi-Dirac factor and $T$ is the transmission probabality.
%where the chemical potential of the
%Fermi function is fixed at the Fermi energy $E_{f}$ provided by the
%DFT code F{\footnotesize IREBALL}.
The relaxation and transmission
calculations are both performed using a double numeric orbital basis
to avoid any discrepancies that may arise using a single numeric
basis \cite{Strange_Jacobsen_JChemPhys08}. Additional details on our
approach can be found in Ref. \cite{Bruque_PRB1}.

%\section{CNT-molecule-CNT model}
We model the experimental CNT-molecule-CNT system in Ref.
\cite{Nuckolls_Sci06} using metallic (7,7) CNT contacts attached to
the $\pi$-cruciform molecule \cite{Klare_Lang04} shown in Fig.
\ref{fig:TE_STRUCT}. The molecule is referred to as molecule {\bf 1}
in Ref. \cite{Nuckolls_Sci06}. To build the structure, we begin by
relaxing the isolated molecule with amide groups attached. We build
the CNT out of optimized unit cells. We then cut the CNT to provide
the closest fit to the molecule with amide linkers attached. We
assume that after the etching step, the CNT contacts are fixed in
their position and the molecular window is governed by the
one-dimensional atomic layer spacing of the CNTs
\cite{Nuckolls_NL07,Nuckolls_Sci06}. We find that 9 unit cells of a
(7,7) CNT are comparable in length to the relaxed molecule plus amide
groups. At the dodecyloxybenzene cross-arm ring ends we attach a
truncated $C_{2}H_{5}$ alkane chain instead of $C_{12}H_{25}$ used
in the experiment since, being insulating, they do not affect the
electronic properties of the conjugated central molecule. The system
is constructed such that each CNT contact is at least 5 unit-cells
in length on either side of the molecule. The CNT at the interface
is passivated with hydrogen atoms to minimize localized surface
states. We find that four CNT unit cells (8 atomic layers) for each
CNT contact are long enough to damp out charge oscillations at the
end layer (where the self-energies are added) that result from the
C-H charge dipoles at the cut interfaces \cite{Bruque_PRB1}.

%\section{Results & Discussion}
We show in Fig. \ref{fig:TE_STRUCT} (B) a planar conformation of
molecule 1 and in (C) a perpendicular conformation rotating only the
cross-arm dodecyloxybenzene rings in each configuration.  Both
systems remain stable when relaxed with the plane of the amide
groups at no more than 14.1 degrees from parallel to the tangential
plane of the CNTs at the point of contact.  This dihedral angle was
found to have minimal effect on transmission up to 15 degrees from
parallel \cite{Qian_CNT_mol_CNT07}. The orientation of the planar
molecule in Fig. \ref{fig:TE_STRUCT} contrasts findings by Ke
\emph{et.al.} modeling a comparable CNT-benzenediamide-CNT system
\cite{Ke_CNT_mol_PRL07}. Ke {\em et al.} found a low $\pi$ orbital
overlap between the (5,5) CNT contacts and molecule where the
molecular plane was above the surface of two (5,5) CNTs.  We find
that the larger 1 nm diameter (7,7) CNT allows the amide linker to
align nearly co-planarly with the CNT surface possibly due to a larger
spacing between H atoms around CNT ends.
The discrepancy could also be affected by the different lengths of the
molecules studied.

In Fig. \ref{fig:TE_STRUCT} (A), we show the calculated transmission
as a function of the difference $E - E_{F}$ for the two structures.
The transmission of the planar molecule has a resonant peak near the
Fermi energy which results in a room temperature (300 K) resistance
of 6.4 M$\Omega$.  The resistance of the perpendicular molecule is
1.6 G$\Omega$. The experimental measured resistance is 5 M$\Omega$
\cite{Nuckolls_Sci06}.

To understand the difference in the transmission curves of the two
molecules in Fig. \ref{fig:TE_STRUCT}, we examine the relaxed
isolated molecular orbitals, shown in Fig. \ref{fig:MOL_ORBITALS}.
Both molecules remain stable during relaxation with the planar
conformation energetically favorable by 1.4 eV. Fig.
\ref{fig:MOL_ORBITALS} shows the LUMO, HOMO, HOMO-1, HOMO-2 and
HOMO-3 (top to bottom) for both molecules with the amide groups
attached at the left and right ends. The difference between the two
molecules is notably the HOMO. In the planar conformation, the HOMO
lies across the molecule perpendicular to the LUMO and the path of
transport. The HOMO in the perpendicular conformation is localized
strongly around the amide linkers and is oriented parallel to the
LUMO.
Because the $\pi$-conjugations extends across both arms of the cross
in the planar molecule, the orbitals are more extended than in the
perpendicular conformation. This spreading of the orbital
wavefunction reduces the HOMO-LUMO gap to 1.69 eV in the planar
conformation from 2.17 eV in the perpendicular conformation. The
HOMO-1 and HOMO-2 states in both molecules are split by a few meV
and are essentially degenerate.

We next compare the covariant spectral functions \cite{Bruque_PRB1}
at each transmission peak shown in Fig. \ref{fig:SPECTRAL} to the orbitals
of the isolated molecules.  The spectral function labels (1-4)
correspond to the labeled transmission peaks in Fig.
\ref{fig:TE_STRUCT}. The broad transmission peak (1) corresponds to
the molecular LUMO in both systems.  The broad transmission peak and
the LUMO that extends across the entire CNT-molecule-CNT
structure indicate that the coupling of the CNT orbital to the
molecular LUMO is strong.  This is consistent with results found by
others \cite{Qian_CNT_mol_CNT07,Ke_CNT_mol_PRL07}. Peak (2) in the
transmission curve results from the coupling of the CNT states to
the HOMO of the planar molecule.
This state is localized on
the cross-arm of the molecule away from the CNTs and is weakly
coupled to the contacts.
The Fano resonance at peak (2)
results from the two parallel paths through the molecule. An
electron can tunnel through the tail of the extended state (1) or it
can tunnel through the localized state (2).
Transmission peak (3) results from the resonant tunneling through
the degenerate HOMO-1 / HOMO-2 states of the planar molecule
localized on the oxygen atoms
linking the dodecyloxybenzene rings to the $C_{2}H_{5}$ alkane chains.
Transmission peak (4) results from coupling to the HOMO-3 orbital of
the planar molecule and the HOMO orbital of the perpendicular
molecule. The spectral functions for peaks (1) and (4) are
qualitatively the same for both molecular conformations, so we only
show the spectral function of the planar molecule. Transmission peak
(3) is also a Fano resonance arising from the two parallel paths
corresponding to the HOMO-1 / HOMO-2 states localized on the
cross-arms and the HOMO-3 state extended across the molecule. This
comparison clearly maps the features in the transmission directly to
the features of the isolated molecular orbitals.

The examinations of the transmission, molecular orbitals and
spectral functions explains the difference in resistance of the two
molecular conformations in Fig. \ref{fig:TE_STRUCT}.  Rotating the the
dodecyloxybenzene rings breaks the conjugation with the rings on the
horizontal axis of the molecule and removes the HOMO localized on
the cross-arms. The HOMO-LUMO gap widens leaving no states near the
Fermi energy to carry current, and the resistance increases by over
2 orders of magnitude.
%Low bias hole current, provided by the HOMO
%state in the planar configuration, is consistent with experimental
%p-type behavior \cite{Nuckolls_NL07,Nuckolls_Sci06}.

The use of conformation change for molecular switching is well
known. Several examples are rotaxane
\cite{FLOOD:SCI:2004,JANG:JACS:2005}, 1,4-bis-phenylethynyl-benzene
\cite{Tomfohr_Sankey_J.Chem.Phys04}, and
2'-amino-4-ethylphenyl-4'-ethylphenyl-5'-nitro-1-benzenethiolate
\cite{REED:APL:2001,REED:CP:2002} which we refer to as the nitro
molecule. In all cases the conformation change of the molecules
alters the molecular orbitals along the path of the electron
transport. In the nitro molecule, the conjugation is broken directly
in the transport path between each contact.  In rotaxane, the
orbital changes from extended to localized along the transmission
path of the molecule. For the cruciform molecule, shown in Fig.
\ref{fig:TE_STRUCT}, the effect of conformation is different. The
rotation of the vertical rings does not affect the
conjugation along the horizontal axis of the molecule, and it does
not localize a previously extended state along the axis of the
molecule. Instead, it removes the HOMO that lies on the cross rings, puts it
back onto the horizontal axis of the molecule but at a lower energy
resulting in exponentially decreased transmission near the Fermi
energy and a several order of magnitude reduction in the
conductance. This is a new twist on the conformation-change paradigm
of molecular switching.

Experimentally it has been suggested that up to two molecular
bridges might be established across the gap \cite{Nuckolls_Sci06}
during the dehydration reaction. While the molecular end groups on
the cut ends of the CNTs are not known, it is reasonable to assume
that the CNT ends remain functionalized with carboxyl groups (COOH),
rather than H atoms, after dehydration reaction
\cite{Nuckolls_Sci06}. To explore these issues, we first add one
additional molecule to our CNT-molecule-CNT system. Fig.
\ref{fig:TE_2MOL_CARBOX} (A) shows a relaxed CNT-molecule-CNT system
where an additional planar molecule is attached parallel to the
original molecule shown in Fig. \ref{fig:TE_STRUCT} (B). The
maximally separated configuration of the two molecules shown here is
known to be energetically favorable. \cite{Guo_CNT_mol_CNT07}. The
two molecule transmission is shown in Fig. \ref{fig:TE_2MOL_CARBOX}
(B) where we observe a doubling of peaks near the Fermi level giving
a calculated resistance of 4.7M$\Omega$. As expected, the peaks are
associated with the same orbitals previously discussed in relation
to Figs. 1-3. The addition of one molecular bridge reduces the resistance,
but not by a factor of two.
The resistance is sensitive to the position of
the HOMO resonant transmission peaks with respect to the Fermi
level, and the two peaks from the two molecules split and shift
compared to the single peak from a single molecule.

Finally, we cut a CNT non-vertically and passivate the side walls
with carboxyl groups.  We attach a single planar molecule shown in
Fig. \ref{fig:TE_2MOL_CARBOX} (C) at the shortest portion of the
molecular gap. The relaxed plane of the amide CONH groups is at no
more than 24.2 degrees from parallel to the tangential plane of the
CNTs at the point of contact.  The transmission is shown in Fig.
\ref{fig:TE_2MOL_CARBOX} (D) where the features remain
qualitatively
comparable to the features of the planar transmission
in Fig. \ref{fig:TE_STRUCT} (A).
Quantitatively, we find a narrowing of each resonant peak.
The LUMO and HOMO-3 peaks shift
slightly deeper into the conductance and valence bands respectively.
The spectral function at each peak again matches the features of the
isolated molecular orbitals shown in Fig. \ref{fig:MOL_ORBITALS}.
The calculated resistance is 40 M$\Omega$.
The increase is the result of the narrowing of the HOMO resonance.
%The resistance for each structure is shown in Table
%\ref{TAB:conduct}.

The transmission curve for the carboxyl passivated system contrasts
to work done by Ren \emph{et. al.} \cite{Guo_CNT_mol_CNT07} where
passivated carboxyl groups where compared to H passivation on
semi-conducting (13,0) CNTs connected by a single diaminobenzene
molecule.  Ren \emph{et. al.} found that the resonant peaks broaden
in the valence and conduction band regions when the side-walls are
passivated with carboxyl.  The narrowing of the valence band
resonances in our case, shown in Fig. \ref{fig:TE_2MOL_CARBOX} (D),
indicate a lower coupling of the molecular states to the continuum
of states in the CNT contacts. The decrease in coupling is partially
due to the increased dihedral angle between the molecule and the
plane of the CNT at the point of contact. The increase of the
dihedral angle is caused by the increase of the steric
hindrance of the carboyxl groups compared to the H atoms.
The CNT sidewall
chemistry, when relaxed, affects the orientation of the molecular
junction due to negatively charged oxygen atoms on the cut surface
of the CNTs repelling the CONH linker oxygen atoms. This repulsion
forces the amide horizontal axis to twist, affecting the $\pi$-bond
overlap between the CNT contacts and the amide linkers.

We note that although the resistance has increased compared to the
hydrogen passivation configuration, the overall transmission curve
remains similar to the original system with the HOMO level near the
Fermi energy. Overall the resistances listed in Table
\ref{TAB:conduct} are close to the experimental measurements,
(excluding the perpendicular conformation). The proximity of the
HOMO state near the Fermi energy provides a weakly-coupled transport
path through the molecule at low bias regardless of the interface
orientation, the number of molecular bridges, or the presence of
carboxyl groups.

In summary, we have found good theoretical agreement with the first
experimental measurement of a CNT-molecule-CNT system.  The
features in the transmission correspond directly to the features of
the isolated molecular orbitals. The rotation of the
dodecyloxybenzene rings of the cross-arm alters the resistance by
over 2 orders of magnitude even though it does not affect the
conjugation along the transport path.
The HOMO lying on the cross-arms of the planar molecule
provides conductance at low bias that is relatively insensitive
to the end groups of the cut CNTs, the cut angle, or the
number of molecular bridges.

\noindent{\bf Acknowledgements}\\
This work is supported by the NSF (ECS-0524501) and the
Semiconductor Research Corporation Focus Center Research Program on
Nano Materials (FENA).

\newpage
\begin{center}
% Conductance Table
\begin{tablehere}
\caption{Resistance of four CNT-molecule-CNT systems studied where
`Planar' indicates the planar conformation of the molecule.
\label{TAB:conduct}} \vspace{10pt}
\begin{tabular}{|c|c|}
  \hline
  System & Conductance (M$\Omega$)\\
  \hline
  Perpendicular  & 1590 \\
  \hline
  Planar & 6.4 \\
  \hline
  Planar / 2 molecules  & 4.7 \\
  \hline
  Planar / carboxyl & 40. \\
  \hline
  Experimental & 5 \\
  \hline
\end{tabular}
\end{tablehere}
\end{center}
\newpage
\noindent {\bf \center Figure Captions} \\

\noindent Fig. \ref{fig:TE_STRUCT}. (A) Calculated transmission of
the CNT-molecule-CNT structures where the solid line (red) is for
the planar dodecyloxybenzene cross-arm conformation and the dashed
line (blue) is for the perpendicular dodecyloxybenzene cross-arm
conformation. (B and C) Relaxed CNT-molecule-CNT planar and
perpendicular structures respectively.
\\
\newline
\noindent Fig. \ref{fig:MOL_ORBITALS}. Calculated molecular orbitals
for the isolated molecule in both the planar (left) and
perpendicular (right) conformations. Amide groups are included at
the left and right ends of each molecule.
\\
\newline
\noindent Fig. \ref{fig:SPECTRAL}. 3D contour plots of the covariant
spectral function corresponding to the resonant transmission peaks
marked in \ref{fig:TE_STRUCT} calculated using the planar
CNT-molecule-CNT structure.
\\
\newline
\noindent Fig. \ref{fig:TE_2MOL_CARBOX}. (A) CNT-molecule-CNT
structure with two planar molecules attached. (B)
Calculated transmission of CNT-molecule-CNT system shown in (A). (C)
CNT-molecule-CNT structure with the CNT side walls passivated with
carboxyl group molecules. (D) Calculated transmission of
CNT-molecule-CNT system shown in (C).
\\

\newpage
\begin{center}
\begin{figurehere}
\includegraphics[width=4.75in]{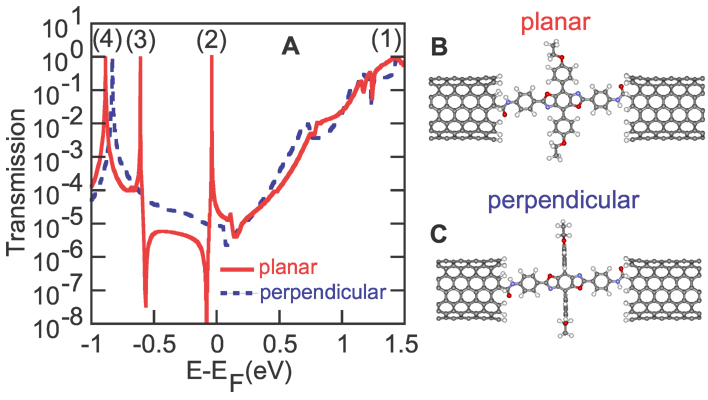}
\caption{\label{fig:TE_STRUCT}}
\end{figurehere}
\newpage
\begin{figurehere}
\includegraphics[width=3.0in]{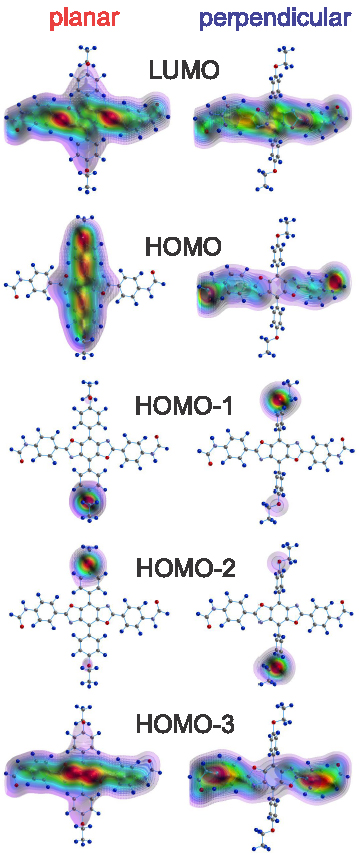}
\caption{\label{fig:MOL_ORBITALS}}
\end{figurehere}
\newpage
\begin{figurehere}
\includegraphics[width=4.75in]{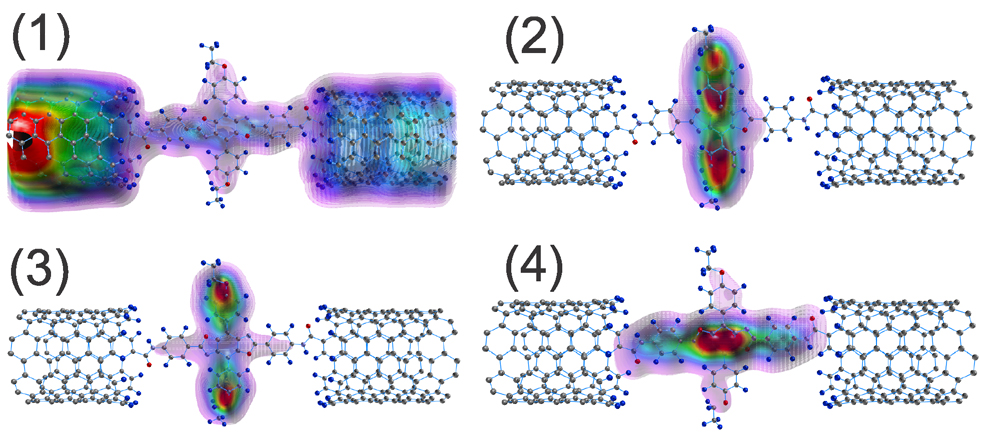}
\caption{\label{fig:SPECTRAL}}
\end{figurehere}
\newpage
\begin{figurehere}
\includegraphics[width=3.0in]{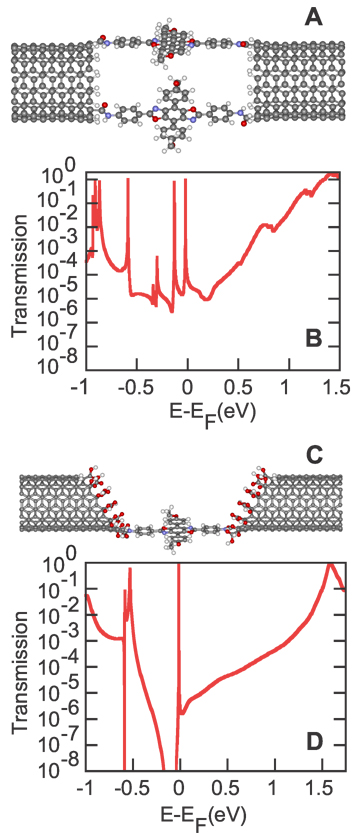}
\caption{\label{fig:TE_2MOL_CARBOX}}
\end{figurehere}
\end{center}
\newpage

%\bibliography{BIBLIOGRAPHY}

\clearpage

\end{document}